\documentclass[%
 reprint,
 amsmath,amssymb,
 aps,
]{revtex4-2}

\usepackage{graphicx}
\usepackage{dcolumn}
\usepackage{bm}
\usepackage{color,xcolor}
\usepackage{lineno}
\usepackage{ulem}

\begin{document}
\preprint{APS/123-QED}

\title{Electric field induced by radial redistribution of the energetic ion pressure in a fusion plasma}

\author{Shaojie Wang}%
 \email{wangsj@ustc.edu.cn}
\affiliation{%
 1. Department of Engineering and Applied Physics, University of Science and Technology of China, Hefei 230026, China, \\
2. Key Laboratory of Frontier Physics in Controlled Nuclear Fusion, Chinese Academy of Sciences, Hefei 230026, China
}%




\date{\today}

\begin{abstract}
It is found by using the gyrokinetic theory that significant radial electric fields, or zonal flows, can be generated by the radial redistribution of energetic ion pressure in a tokamak fusion device. Trapped energetic ions are more effective to generate the radial electric field than the isotropic energetic ions. This suggests that the energetic $\alpha$ particles produced by DT fusion may induce significant radial electric field and thus help to improve the core plasma confinement in a fusion reactor.   

\end{abstract}

\maketitle



Recent experiments have shown that the core confinement of magnetic fusion plasma is significantly improved by energetic ions (EIs) \cite{MazziNP22,HanNature22,ZhangNF22,DengPOP22,GarciaNC24}. EIs can be preduced by auxiliary heating through injection of the neutral beams or the ion-cyclotron-range of frequency waves in the present devices, and by the DT fusion reactions in a fusion reactor, such as ITER \cite{ITERNF99}; this implies that the fusion produced energetic $\alpha$ particles may be benificial to the core plasma confinement, which is crucially important for a future DT fusion reactor \cite{DoyleNF07}. Interaction between EIs and turbulence is of significant interest not only in magnetic fusion when facing the burning plasmas \cite{SalewskiNF25,NaNF25}, but also in many orther fields, such as (but not limited to) the space plasma physics \cite{LemoinePRL22} and the astrophysical science \cite{BykovSSR19,TrottaAJ20}.

The equilibrium EI distribution indicates that its low energy part can resonate with the low-frequency drift wave, such as the ion-temperature-gradient (ITG) mode \cite{RomanelliPoP93}, which usually determines the ion anomalous transport in a magnetic fusion device; when the EI effective temperature is a few times of the thermal ion temperature, these resonant particles linearly stabilizes the ITG mode  \cite{SienaPRL20, SienaNF19}; however, when the EI effective temperature is much higher than the thermal ion temperature, which is the case of the energetic $\alpha$ particles produced in a DT fusion reactor, the resonant EI's linear stabilizing effect disappears, with the dilution effect left \cite{SienaNF19}. 

Suppression of the ITG turbulence by ZFs \cite{LinScience98,RosenbluthPRL98} or radial electric field shearing \cite{BiglaryPFB90} has been well established. The transport theory indicates that ZFs can be nonlinearly driven by the poloidal Reynolds Stress (RS) \cite{DiamondPRL94}, and by the Radial Redistribution (RR) of thermal ions \cite{WangPoP17,ZhangNF20}; the wave-wave interaction theory indicates that ZFs can be nonlinearly driven by the modulational instability \cite{ChenPoP00}. 
Recent nonlinear global gyrokinetic (GK) simuilations have confirmed the effects of RS and RR of thermal ion pressure \cite{WangPRE22,WangNF25} and their importance \cite{WangPRL24} in the core confinement improvement \cite{ConnorNF04,IdaPPCF18}. Note that the RR of thermal ion pressure denotes the (mesoscale) transport of thermal ion energy and the RS denotes the (mesoscale) transport of thermal ion momentum \cite{WangPoP17,ZhangNF20}. 
Although the EI-induced geodesic acoustic mode \cite{FuPRL08}, which is also known as the high-frequency brahch of ZFs, may be less effective in suppressing turbulence than ZFs, EIs can drive ZFs through exciting the Alfvenic modes \cite{ChenPRL12,QiuPoP16,QiuNF16}. The wave-wave interaction theory predicts that the strength of this beam-driven ZFs is determined by the fluctuation intensity of the excited Alfvenic eigenmode \cite{QiuPoP16,QiuNF16}, instead of the macroscopic physical quantities, such as RS \cite{DiamondPRL94} and pressure RR \cite{WangPoP17} which are routinely measured in experiments; particularly, it depends on the instability of the Alfvenic eigenmode, which usually needs the EI pressure exceed a critical value.  

In this paper, we propose the GK theory of radial electric field induced by RR of the EI pressure. It is found that significant ZFs, or the radial electric field, can be driven by RR of the EI pressure; trapped EIs are more effective than isotropic EIs in driving ZFs. In particular, the proposed theory predicts a significant level of the mean radial electric field, $\sim 30\rm{kV/m}$, produced in the core plasma of ITER by the DT fusion itself. 

Consider ZFs with EIs in a large-aspect ratio tokamak, $\epsilon=r/R\ll 1$, with $r$ and $R$ the minor and major radii of the torous, respectively.  
We begin with the ensemble-averaged nonlinear GK equation \cite{BrizardRMP07,WangPoP17,ZhangNF20},  
\begin{equation}\label{eq:GKE0}
\partial_t \delta f_{\rm{s}} + \mathcal{L}\delta f_{\rm{s}}
= e_{\rm{s}} J_0 \delta\phi \partial_{\mathcal{E}} F_{\rm{s}}+ J_0 \partial_t \delta \mathcal{R}_{\rm{s}}F_{\rm{s}},
\end{equation}
and the quasi-neutrality equation (QNE),
\begin{equation}\label{QNE0}
\nabla\cdot\left(\varepsilon_{\rm{c}}\nabla_{\perp}\delta\phi\right)+\sum_{\rm{s}}e_{\rm{s}} \langle  \left(1,~J_0\delta f_{\rm{s}}\right) \rangle_{\rm{r}}=0.
\end{equation}
Here $\delta f_{\rm{s}}\left(r,\theta,\mu,\mathcal{E}\right)$ is the perturbed gyrocenter distribution of particle species $\rm{s}$, with $\rm{s}=\rm{e},~\rm{i},~\rm{h}$ denotes the thermal electrons, the thermal ions and the EIs, respectively. $\mu$ and $\mathcal{E}$ are the magnetic moment and the particle energy, respectively. $\delta\phi=\delta\phi e^{ik_{r}r}$ is the electrostatic potentail perturbation of ZFs.
$\mathcal{L}=\dot{\bm{X}}_0\cdot \nabla$, and $\dot{\bm{X}}_0$ is the gyrocenter velocity in the equilibrium field. 
$\varepsilon_{\rm{c}}=\sum_{\rm{s}}\varepsilon_{\rm{c},\rm{s}}$, $\varepsilon_{\rm{c},\rm{s}}=n_{\rm{s}} m_{\rm{s}}/B^2$. The equilibrium magnetic field is $\bm{B}=B_{\zeta}\bm{e}_{\zeta}+B_{\theta}\bm{e}_{\theta}$, with $B_{\theta}\ll B_{\zeta}$. $r$, $\theta$ and $\zeta$ are minor radius, poloidal angle and toroidal angle respectively. $e_{\rm{s}}$ and $m_{\rm{s}}$ are particle charge and mass, respectively. 
$n_{\rm{s}}=\left(1,~F_{\rm{s}} \right)$ is the equilibrium particle dentsity. Note that $J_0(k_{r}\rho_{\rm{s}})=1-\frac{1}{4}k_{r}^2 \rho_{\rm{s}}^2$ denotes the gyro-average, with $\rho_{\rm{s}}=m_{\rm{s}}v_{\perp}/e_{\rm{s}}B$ the Larmor radius and $v_{\perp}$ the perpendicular velocity. $\langle\cdot\rangle_{\rm{r}}$ denotes averaging over the magnetic-flux-surface labelled by $r$. 
$\left(f,~g\right)\equiv \int \text{d}^3 \bm{v} f g $. The equilibrium distribution $F_{e,i}$ is assumed Maxwellian, while $F_{\rm{h}}$ is assumed isotropic. $(m_{\rm{s}}v_{\|}^2,~F_{\rm{s}})=n_{\rm{s}}T_{\rm{s}}=(\mu B,~F_{\rm{s}})$, with $v_{\|}$ the parallel velocity. $P_{\rm{s}}=n_{\rm{s}}T_{\rm{s}}$. 

The source term $\delta \mathcal{R}_{\rm{s}}=\delta \mathcal{R}_{\rm{s}} e^{ik_r r}$ represents the RR of density ($\delta n_{\rm{s}}$), energy ($1.5\delta P_{\rm{s}}$), and toroidal momentum ($n_{\rm{s}}\delta \mathcal{U}_{\zeta,\rm{s}}$), due to nonlinear transport or external injection; $\partial_t\delta\mathcal{R}_{\rm{s}}F_{\rm{s}}$ equals to the external source minus the divergence of the (mesoscale) phase-space transport flux \cite{WangPoP17}. 

Making the Legendre expansion in terms of $v_{\|}/v$, one finds \cite{WangPoP17} $\delta \mathcal{R}_{\rm{s}}=\delta \mathcal{R}_{\rm{s},0}\left(\mathcal{E}\right)+\delta \mathcal{R}_{\rm{s},1}$, with $\delta \mathcal{R}_{\rm{s},1}=m_{\rm{s}}\delta \mathcal{U}_{\zeta,\rm{s}} v_{\|}/T_{\rm{s}}$. Note that \cite{RosenbluthPRL98} $\delta \mathcal{R}_{\text{s}, l}\sim\mathcal {O}\left(k_{r}\rho_{\rm{s}}\right)^l$. The RR of particle density, pressure, and toroidal momentum are respectively given by $\delta n_{\rm{s}}=(1,~ \delta\mathcal{R}_{\rm{s},0}F_{\rm{s}})$, $\delta P_{\rm{s}}=(2\mathcal{E}/3,~ \delta\mathcal{R}_{\rm{s},0}F_{\rm{s}})$, and $n_{\rm{s}}\delta \mathcal{U}_{\zeta,\rm{s}}=(v_{\|},~ \delta\mathcal{R}_{\rm{s},1}F_{\rm{s}})$. This $\delta\mathcal{R}_{\rm{s},1}$ term describes essentially the RR of parallel momentum. 
The RR of the perpendicular momentum should be discussed here; it denotes the non-ambipolar tranport term or the poloidal RS in an electrostatic turbulence \cite{ZhangNF20}, and it is $\sim \mathcal{O}\left(k_{r} \rho_{\rm{s}}\right)^2$. To separate it from $\delta\mathcal{R}_{\rm{s},0}$ and write it as $\delta\mathcal{R}_{\rm{s},2}$, we find
\begin{equation}\label{eq:nonamb}
\delta \mathcal{R}_{\rm{s},2}=ik_{r}\epsilon_{\rm{c},\rm{s}}\frac{1}{n_{\rm{s}} e_{\rm{s}}}\left(\delta \mathcal{U}_{\zeta,\rm{s}}B_{\theta}-\delta \mathcal{U}_{\theta,\rm{s}}B_{\zeta}\right).
\end{equation}
This can be easily understood by noting that the GK polarization charge associated with $n_{\rm{s}}\delta\bm{\mathcal{U}}_{\rm{s}}$, which denotes the momentum RR induced by either the turbulent transport or the external injection, can be written as $-\nabla\cdot\delta\bm{\mathcal{P}}_{\rm{s}}$, with $\delta\bm{\mathcal{P}}_{\rm{s}}=\epsilon_{\rm{c},\rm{s}}\delta\bm{\mathcal{U}}_{\rm{s}}\times\bm{B}$. The $\delta \mathcal{U}_{\theta,\rm{s}}$ term denotes the contribution from the poloidal RS \cite{DiamondPRL94}, $\pi_{r,\theta}$; by letting $n_{\rm{s}}\partial_t \delta \mathcal{U}_{\theta,\rm{s}}=-ik_{r}\pi_{r,\theta}$, it recovers the previous nonlinear GK theory \cite{ZhangNF20}; clearly, the poloidal RS denotes the RR of poloidal momentum. With the non-ambipolar term moved out of $\delta \mathcal{R}_{\rm{s},0}$, $\delta n_{\rm{s}}$ contained in $\delta \mathcal{R}_{\rm{s},0}$, is clearly ambipolar \cite{ZhangNF20}, which implies 
\begin{equation}\label{eq:QN}
\sum_{\rm{s}} e_{\rm{s}}(1,~ \delta\mathcal{R}_{\rm{s},0}F_{\rm{s}})=0.
\end{equation}

To solve Eq. (\ref{eq:GKE0}), we introce the transform $\delta f_{\rm{s}}=\delta g_{\rm{s}}+ e_{\rm{s}}J_0\delta\phi \partial_{\mathcal{E}}F_{\rm{s}}$, to find
\begin{equation}\label{eq:GKE}
\partial_t \delta g_{\rm{s}} + \mathcal{L}\delta g_{\rm{s}}
= -e_{\rm{s}} J_0 \partial_t \delta\phi \partial_{\mathcal{E}} F_{\rm{s}}+ J_0 \partial_t \delta \mathcal{R}_{\rm{s}}F_{\rm{s}}.
\end{equation}
Following Ref. \onlinecite{RosenbluthPRL98}, the solution of $\delta g_{\rm{s}}$ is found by using the condition $\partial_t\ll \mathcal{L}$ for ZFs. Writing $\delta g_{\rm{s}}=\delta g_{\rm{s}} e^{ik_r\overline{r}}$, one finds  
\begin{equation}
\delta g_{\rm{s}}=-e_{\rm{s}}\mathcal{J}_0J_0\delta \phi \partial_{\mathcal{E}}F_{\rm{s}}+\mathcal{J}_0J_0 \delta \mathcal{R}_{\rm{s}}F_{\rm{s}},
\end{equation}
with $\mathcal{J}_0$ the orbital averaging operator. $\mathcal{J}_{0} A\equiv \overline{A}\equiv \frac{1}{\tau_{\theta}}\oint\rm{d}\theta \frac{1}{\dot{\theta}}A$, with $\tau_{\theta}$ the bounce/transit time ($\tau_b$/$\tau_t$) for trapped/passing particles. 

Note that the flux-surface average of a velocity integral equals the velocity integral of the orbital average \cite{HazeltineBook92},
\begin{equation}
\langle \left(1,~ A\right) \rangle_{\rm{r}}=\left(1,~ \mathcal{J}_0 A\right), 
\end{equation}
\begin{equation}
\int \text{d}^3 \bm{v}=\int_0^{\infty}4\pi v^2 \text{d}v\left(\int_0^{\lambda_{\rm{c}}} \text{d}\lambda \sum_{\sigma}+\int_{\lambda_{\rm{c}}}^{\lambda_{\rm{m}}} \text{d}\lambda  \right)\frac{1}{\mathcal{N}}, 
\end{equation}
with $\lambda=\mu B_0/\mathcal{E}$, $\sigma=sign(v_{\|})$. $1/\mathcal{N}\left(\lambda,r\right)=\tau_{\theta}v/8\pi qR$. $q$ is the safety factor. $\lambda_{\rm{c}}=1-\epsilon$ is the trapping-passing boundary, and $\lambda_{\rm{m}}=1+\epsilon$. 
The QNE is thus reduced to
\begin{equation}
\begin{split}
&-k_{r}^2 \varepsilon_{\rm{c}} \delta\phi-\sum_{\rm{s}}e^2_{\rm{s}}\left(J_0^2 \left(1-\mathcal{J}_0^2\right)\delta\phi,~-\partial_{\mathcal{E}} F_{\rm{s}}\right)\\
&=\sum_{\rm{s}}e_{\rm{s}}\left(J_0^2 \mathcal{J}_0^2 \delta \mathcal{R}_{\rm{s}},~F_{\rm{s}}\right).
\end{split}
\end{equation}

Set $\left(\Delta r\right)_{\rm{s}}\equiv r-\overline{r}=\frac{m_{\rm{s}}\left(v_{\|}-\overline{v_{\|}}\right)} {e_{\rm{s}}B_{\theta}}$. 
Taylor expansion to $\mathcal{O}\left(k_{r}\rho_{\rm{s}}\right)^2$, yields 
$\mathcal{J}_0\delta\phi=\left(1-\frac{1}{2}k_{r}^2 \overline{(\Delta r)_{\rm{s}}^2}\right)\delta\phi$, 
\begin{equation*}
\mathcal{J}_0\delta\mathcal{R}_{\rm{s},0}=\left(1-\frac{1}{2}k_{r}^2 \overline{\left(\Delta r\right)_{\rm{s}}^2}\right)\delta\mathcal{R}_{\rm{s},0},
\end{equation*}
\begin{equation*}
\mathcal{J}_0\delta\mathcal{R}_{\rm{s},1}=ik_{r} \overline{\left(\Delta r\right)_{\rm{s}}^2}\frac{e_{\rm{s}} }{T_{\rm{s}}}\delta \mathcal{U}_{\zeta,\rm{s}} B_{\theta},
\end{equation*}
\begin{equation*}
\mathcal{J}_0\delta\mathcal{R}_{\rm{s},2}=ik_{r}\epsilon_{\rm{c},\rm{s}}\frac{1}{n_{\rm{s}} e_{\rm{s}}}\left(\delta \mathcal{U}_{\zeta,\rm{s}} B_{\theta}-\delta \mathcal{U}_{\theta,\rm{s}} B_{\zeta}\right). 
\end{equation*}

The QNE is further reduced to
\begin{equation}\label{eq:QNE1}
\begin{split}
-&k_{r}^2 \varepsilon_{\rm{c}} \delta\phi-k_{r}^2\sum_{\rm{s}}\left(n_{\rm{s}} e^2_{\rm{s}}\overline{(\Delta r)_{\rm{s}}^2} ,~ -\frac{1}{n_{\rm{s}}}\partial_{\mathcal{E}}F_{\rm{s}}\right)\delta\phi\\
=&-k_{r}^2\sum_{\rm{s}}\left(\frac{1}{2}n_{\rm{s}} e^2_{\rm{s}}\rho_{\rm{s}}^2 +n_{\rm{s}} e^2_{\rm{s}}\overline{(\Delta r)_{\rm{s}}^2} ,~\frac{1}{n_{\rm{s}} e_{\rm{s}}}\delta \mathcal{R}_{\rm{s},0}F_{\rm{s}}\right)\\
&+ik_{r} \sum_{\rm{s}}\left(n_{\rm{s}} e^2_{\rm{s}}\overline{\left(\Delta r\right)_{\rm{s}}^2},~\frac{1}{n_{\rm{s}} T_{\rm{s}}}F_{\rm{s}} \right)\delta \mathcal{U}_{\zeta,\rm{s}} B_{\theta}\\
&+ik_{r} \sum_{\rm{s}}\epsilon_{\rm{c},\rm{s}}\left(\delta \mathcal{U}_{\zeta,\rm{s}} B_{\theta}-\delta \mathcal{U}_{\theta,\rm{s}} B_{\zeta}\right), 
\end{split}
\end{equation}
where the last term on the right-hand side denotes the non-ambipolar transport, and we have used Eq. (\ref{eq:QN}). 

Following Ref. \onlinecite{WangPoP17}, we have
\begin{equation}
\frac{1}{\mathcal{N}}n_{\rm{s}}e_{\rm{s}}^2\overline{(\Delta r)_{\rm{s}}^2} =\frac{n_{\rm{s}}m_{\rm{s}}}{B_{\theta}^2}\mathcal{E}\frac{2\sqrt{2\epsilon}}{\pi}\mathcal{K}(\kappa^2),
\end{equation}
with $\mathcal{K}(\kappa^2)=\mathcal{K}_b(\kappa^2)=
\left[E(\kappa^2)+(\kappa^2-1)K(\kappa^2)\right]$, for trapped particles, and 
$\mathcal{K}(\kappa^2)=\mathcal{K}_t(\kappa^2)=\frac{1}{2}\kappa
\left[E(1/\kappa^2)-\frac{(\pi/2)^2}{K(1/\kappa^2)}\right]$, for passing particles. 
Here $E$ and $K$ are complete elliptic integrals of the first and second kind, respectively. $\kappa^2=\left(1+\epsilon-\lambda\right)/2\epsilon$. For trapped particles, $\kappa^2=\sin^2\left(\theta_b/2\right)$, with $\theta_b$ the bounce angle. Note that
$\tau_{b}=\frac{4\sqrt{2}qR}{\sqrt{\epsilon}v}K\left(\kappa^2\right)$, and
$\tau_{t}=\frac{2\sqrt{2}qR}{\sqrt{\epsilon}v}\frac{1}{\kappa}K\left(1/\kappa^2\right)$.

Therefore, the $\lambda$-integral in Eq. (\ref{eq:QNE1}) can be straightforwardly evaluated for the cases with isotropic $F_{\rm{s}}$. 
\begin{equation}\label{eq:deltar2}
\begin{split}
&\left( \sum_{\sigma}\int_0^{\lambda_{\rm{c}}} \rm{d}\lambda+\int_{\lambda_{\rm{c}}}^{\lambda_{\rm{m}}} \rm{d}\lambda  \right)\frac{1}{\mathcal{N}} n_{\rm{s}} e_{\rm{s}}^2\overline{(\Delta r)_{\rm{s}}^2}\\
&=\frac{4\sqrt{2}q^2}{\pi\epsilon^{1/2}}\epsilon_{\rm{c},\rm{s}}\cdot\mathcal{E}\left(2\int_1^{\frac{1+\epsilon}{2\epsilon}}\rm{d}\kappa^2 +\int_0^1 \rm{d}\kappa^2 \right) \mathcal{K}(\kappa^2)\\
&=1.6q^2\epsilon^{-1/2}\epsilon_{\rm{c},\rm{s}}\cdot\frac{2}{3}\mathcal{E},
\end{split}
\end{equation}
\begin{equation}\label{eq:rho2}
\left( \sum_{\sigma} \int_0^{\lambda_{\rm{c}}} \rm{d}\lambda+\int_{\lambda_{\rm{c}}}^{\lambda_{\rm{m}}} \rm{d}\lambda  \right)\frac{1}{\mathcal{N}}n_{\rm{s}} e_{\rm{s}}^2\frac{\rho^2}{2}  =\epsilon_{\rm{c},\rm{s}}\cdot\frac{2}{3}\mathcal{E}.
\end{equation}

After the $\lambda$-integral in Eq. (\ref{eq:QNE1}) is evaluated, one is ready to evaluate the $v$-integral to find the dispersion relation. 
\begin{equation}\label{eq:dispersion}
\begin{split}
&-ik_{r}  \varepsilon_{\rm{r}}\epsilon_{\rm{c}}\delta\phi
-\varepsilon_{\rm{r}} \sum_{\rm{s}}\epsilon_{\rm{c},\rm{s}}\frac{ik_{r} \delta P_{\rm{s}}}{n_{\rm{s}} e_{\rm{s}}}\\
&=\varepsilon_{\rm{r}}\sum_{\rm{s}}\epsilon_{\rm{c},\rm{s}}\delta \mathcal{U}_{\zeta,\rm{s}}B_{\theta}
-\sum_{\rm{s}}\epsilon_{\rm{c},\rm{s}}\delta \mathcal{U}_{\theta,\rm{s}}B_{\zeta},
\end{split}
\end{equation}
with $\epsilon_{\rm{r}}=1+1.6q^2/\sqrt{\epsilon}$ the well-konwn neoclassical polarization factor \cite{RosenbluthPRL98}. In finding Eq. (\ref{eq:dispersion}), we have assumed that the EIs' equilibrium is either the Maxwellian or the isotropic slowing-down distribution when evaluating the $\partial_{\mathcal{E}}F_{\rm{h}}$ term in Eq. (\ref{eq:QNE1}). 
Note that $n_{\rm{h}}\ll n_{\rm{i}}$ for typical cases, therefore, the above result is insensitive to the details of EIs' isotropic distribution. 

For a Maxwellian equilibrium distribution of EIs,  
\begin{equation}
F_{\rm{h}}=F_{\rm{h},\rm{M}}=\frac{n_{\rm{h}}}{(2\pi T_{\rm{h}}/m_{\rm{h}})^{3/2}}e^{-\mathcal{E}/T_{\rm{h}}},
\end{equation} 
the Laguerre expansion of the source term is given by \cite{WangPoP17}
\begin{equation}
\delta \mathcal{R}_{\rm{h},0}=\frac{\delta n_{\rm{h}}}{n_{\rm{h}}}+\frac{\delta T_{\rm{h}}}{T_{\rm{h}}}\left(\frac{\mathcal{E}}{T_{\rm{h}}}-\frac{3}{2}\right).
\end{equation}

For an isotropic slowing-down equilibrium distribution EIs, 
\begin{equation}
F_{\rm{h}}=F_{\rm{h},0}\equiv\frac{3(m_{\rm{h}}/2)^{3/2}}{4\pi\log\frac{\mathcal{E}_{\rm{b}}^{3/2}+\mathcal{E}_{\rm{c}}^{3/2}}{\mathcal{E}_{\rm{c}}^{3/2}}}\cdot\frac{n_{\rm{h}}}{\mathcal{E}^{3/2}+\mathcal{E}_{\rm{c}}^{3/2}},
\end{equation}
with $\mathcal{E}_{\rm{b}}$ the birth (maximum) energy of EI, and $\mathcal{E}_{\rm{c}}$ the critical energy above which the slowing-down of EI is dominated by electrons. 
Note that $\left(1,~F_{\rm{h}}\right)=n_{\rm{h}}$, $\left(\mathcal{E},~ F_{\rm{h}}\right) \equiv (3/2)P_{\rm{h}}\equiv (3/2)n_{\rm{h}}T_{\rm{h}}$. 
$T_{\rm{h}}=(2/3)\mathcal{E}_{\rm{b}}/\log\left(\mathcal{E}_{\rm{b}}/\mathcal{E}_{\rm{c}}\right)$. We have ignored higher order terms by assuming $\mathcal{E}_{\rm{c}}/\mathcal{E}_b\ll 1$.
For this case, the generalized Laguerre expansion of the source term is given by 
\begin{equation}
\delta \mathcal{R}_{\rm{h},0}=\frac{\delta n_{\rm{h}}}{n_{\rm{h}}}+\frac{\delta T_{\rm{h}}}{T_{\rm{h}}}\left(\frac{\mathcal{E}}{T_{\rm{h}}}-\frac{3}{2}\right)\frac{4/3}{\log\left(\mathcal{E}_{\rm{b}}/\mathcal{E}_{\rm{c}}\right)-2}.
\end{equation}

In neglecting the EI effects, the present theory [Eq. (\ref{eq:dispersion})] recovers the previous theories on the RR effect of thermal ions \cite{WangPoP17,ZhangNF20}, which has been confirmed by recent nonlinear GK simulations \cite{WangPRE22,WangNF25}. The RR theory presented here predicts that the growth rate of ZFs is twice of the linear pumping mode in the quasilinear stage by noting the fact that the growth rate of the quasilinear transport fluxes is simply twice of the linear growth rate, which has been clearly demonstrated in the nonlinear global GK simulation \cite{WangPRE22} for ZFs self-generated by the ITG turbulence and further confirmed in a  recent GK simulation \cite{RiemannPRL25}; this behavior of the growth rates is similar to the wave-wave interaction theory of the beam-driven Alfvenic-eigenmode-induced ZFs \cite{QiuPoP16, QiuNF16}. In contrast to the wave-wave interaction theory \cite{QiuPoP16, QiuNF16}, which predicts that the strength of the beam-driven ZFs is determined by the fluctuation intensity of the EI-dirven Alfvenic eigenmodes, the present RR theory, Eq. (\ref{eq:dispersion}), predicts that the ZF, or the mean radial electric field, is determined by the RR of EIs, which can be routinely measured in experiments. More importantly, the present transport theory is not sensitive to the details of fluctuations.

The last term on the right-hand side of Eq. (\ref{eq:dispersion}) is the non-ambipolar term \cite{RosenbluthPRL98,WangPoP17}, which represents the RS effect \cite{ZhangNF20} and has been confirmed in recent nonlinear GK simulations \cite{WangPRE22, WangNF25}; the RS term is shielded by the neoclassical polarization effect on the time scale longer than the ion bounce time \cite{WangNF25}. Recent nonlinear GK simulation \cite{WangPRL24} has shown that both the thermal ion pressure RR and the poloidal RS are important in the dynamics of internal transport barrier formation. For mathematical simplicity, we shall ignore this RS term in the following. 

Following Eq. (\ref{eq:dispersion}), the RR-driven radial electric field ($\delta E_{\rm{r}}=ik_{r} \delta \phi$) is given by
\begin{equation}\label{eq:Er0}
\delta E_{\rm{r}}=\frac{1}{\rho_{\rm{M}}}\sum_{\rm{s}}\left(\frac{m_{\rm{s}}}{e_{\rm{s}}}\delta P'_{\rm{s}}+\rho_{M,s}\delta \mathcal{U}_{\zeta,\rm{s}}B_{\theta}\right), 
\end{equation}
with $\rho_{\rm{M}}=\rho_{\rm{M},\rm{i}}+\rho_{\rm{M},\rm{h}}$, $\rho_{M,s}=n_{\rm{s}} m_{\rm{s}}$, $\delta P'=\rm{d}\delta P/\rm{d}r$. Note that for typical cases with $n_{\rm{h}}\ll n_{\rm{i}}$, $\rho_{\rm{M},\rm{h}}\ll \rho_{\rm{M},\rm{i}}$; therefore, the effects of RR of EI's momentum can be ignored, and it is EI's pressure RR that is important in driving the radial electric field. 

To numerically confirm the main result of this paper, we have carried out a gereralized Rosenbluth-Hinton test by using the NLT code \cite{YeJCP16, XuPoP17}, which is based on the numerical Lie-transform method \cite{WangPoP12, WangPRE13, WangPoP13}. The initial perturbation is set up with $\delta n_{\rm{i}}e_{\rm{i}}+\delta n_{\rm{h}}e_{\rm{h}}=0$ and an isotropic slowing-down distribution of EIs; the numerical results, which agree well with Eq. (\ref{eq:Er0}), indicate that the radial electric field is mainly driven by RR of the EI pressure. 

We further consider the trapped EIs with their equilibrium given by
\begin{equation}
F_{\rm{t}}=\frac{\pi}{\sqrt{2\epsilon}K\left(\kappa_b^2\right)}\delta\left(\kappa^2-\kappa_b^2\right)F_{\rm{h},0}. 
\end{equation}
Here, the subscript $\rm{t}$ denotes the trapped EIs. Note that $\left(1,~F_{\rm{t}}\right)=n_{\rm{t}}$, $\left(\mathcal{E},~ F_{\rm{t}}\right)\equiv P_{\rm{t}}\equiv n_{\rm{t}}T_{\rm{t}}$. 
$T_{\rm{t}}=\mathcal{E}_{\rm{b}}/\log\left(\mathcal{E}_{\rm{b}}/\mathcal{E}_{\rm{c}}\right)$. 
For the case of trapped EIs, we shall assume $\delta \mathcal{R}_{1,s}=0$, ignoring the RR of toroidal momentum.  
$\delta n_{\rm{t}}=(1,~\delta \mathcal{R}_{0,\rm{t}}F_{\rm{t}})$, $\delta P_{\rm{t}}=(\mathcal{E},~\delta \mathcal{R}_{0,\rm{t}}F_{\rm{t}})$. 

For the case of trapped EIs, the procedure to evaluate the $\lambda$-integrals in Eq. (\ref{eq:QNE1}) is similar to Eqs. (\ref{eq:deltar2}, \ref{eq:rho2}) but the factor $F_{\rm{t}}/F_{\rm{h},0}$ should be included in the integrand, and the results are respectively given by 
\begin{equation*}
4\frac{\mathcal{K}_b(\kappa_b^2)}{K\left(\kappa_b^2\right)}\cdot\frac{q^2}{ \epsilon}\epsilon_{\rm{c},\rm{t}}\mathcal{E},~ \epsilon_{\rm{c},\rm{t}} \mathcal{E}.
\end{equation*}


The dispersion relation of ZFs driven by trapped EIs is given by 
\begin{equation}\label{eq:dispersionT}
k_{r}^2\sum_{\rm{s}}\epsilon_{\rm{c},\rm{s}}\left(\varepsilon_{r,s}^{\phi}\delta\phi+\varepsilon_{r,s}^{P}\frac{\delta P_{\rm{s}}}{n_{\rm{s}} e_{\rm{s}}}\right)=0.
\end{equation}
Here $\varepsilon_{r,i}^{\phi}=\varepsilon_{\rm{r}}=\varepsilon_{r,i}^{P}$, and
\begin{equation}
\varepsilon_{\rm{r},\rm{t}}^{\phi}=1+6\frac{\mathcal{K}_b(\kappa_b^2)}{K\left(\kappa_b^2\right)}\cdot\frac{q^2}{ \epsilon}.
\end{equation}
\begin{equation}
\varepsilon_{\rm{r},\rm{t}}^{P}=1+4\frac{\mathcal{K}_b(\kappa_b^2)}{K\left(\kappa_b^2\right)}\cdot\frac{q^2}{ \epsilon}.
\end{equation}


Following Eq. (\ref{eq:dispersionT}), ZFs driven by the pressure RR of trapped EIs is given by
\begin{equation}\label{eq:ErT}
\delta E_{\rm{r}}=\frac{1}{\rho_{\rm{M},\rm{eff}}}\left(\frac{m_{\rm{i}}}{e_{\rm{i}}}\delta P'_{\rm{i}}+\frac{\epsilon_{\rm{r},\rm{t}}^P}{\epsilon_{\rm{r}}}\frac{m_{\rm{t}}}{e_{\rm{t}}}\delta P'_{\rm{t}}\right),
\end{equation}
with $\rho_{\rm{M},\rm{eff}}=\rho_{\rm{M},\rm{i}}+\rho_{\rm{M},\rm{t}}\varepsilon_{\rm{r},\rm{t}}^{\phi}/\varepsilon_{\rm{r}}$. Note that for typical cases with $n_{\rm{t}}\ll n_{\rm{i}}$, $\rho_{\rm{M},\rm{eff}}\approx n_{\rm{i}}m_{\rm{i}}$. 

For $\kappa_b^2=1/2$, $4\frac{\mathcal{K}_b(\kappa_b^2)}{K\left(\kappa_b^2\right)}=0.91$; this result agrees well with the model of trapped EIs uniformly distributed in the trapping region. 
\begin{equation}
\varepsilon_{\rm{r},\rm{t}}^{P}/\varepsilon_{\rm{r}}\sim\epsilon^{-1/2}.
\end{equation} 
This enhancement factor for the trapped EIs can be easily understood, since the neoclassical polarization is mainly contributed by the trapped particles. Clearly, the trapped EIs are more effective in driving ZFs through the pressure RR than the isotropic EIs. This suggests that the RR of trapped EIs induced by the fishbone instability \cite{McGuirePRL83, ChenPRL84} may generate significant ZFs, which is consistent with recent experimental observations on the internal transport barrier formation triggered by the fishbone mode \cite{BrochardPRL24}. 

The present theory may be used to predict the ZFs or the mean radila electric fields driven by the pressure RR of EIs, the fusion produced $\alpha$ particles in a tokamak fusion reactor. For the typical parameters of ITER \cite{ITERNF99}, consider an isotropic slowing-down distribution of energetic $\alpha$ particles produced by DT fusion. $n_{\rm{h}}\ll n_{\rm{i}}$, but $P'_{\rm{h}}\sim P'_{\rm{i}}$ in the core plasma. Assuming $\delta P'_{\rm{s}}\propto P'_{\rm{s}}$ in a turbulence, one finds $\delta P'_{\rm{h}}\sim \delta P'_{\rm{i}}$, therefore, one predicts significant ZFs driven by the pressure RR of energetic $\alpha$ particles by using Eq. (\ref{eq:Er0}). 

Note the fact that the RR of EIs can be induced by the external injection or the DT fusion reaction, in addition to the nonlinear transport.
It is interesting to consider the mean radial electric field produced by the DT fusion in ITER. The DT fusion annihalates the thermal fuel ions but produces the energetic $\alpha$ particles in the system. The RR of particle density and thus the RR of the fuel ion pressure by fusion reactions are negligible, however, the pressure RR of the energetic $\alpha$ particles by fusion reactions may be significant. For the typical parameters of ITER \cite{ITERNF99}, the DT fusion reactions convert a trace amount of fuel ions into energetic $\alpha$ particles in the core, with a change of the EIs pressure gradient, $\Delta P'_{\rm{h}}\sim P'_{\rm{h}}\sim P'_{\rm{i}}$. Therefore, Eq. (\ref{eq:Er0}) predicts that the DT fusion reactions in ITER generate a significant increment of the mean radial electric field, $\Delta E_{\rm{r}} =0.8 \Delta P'_{\rm{h}}/n_{\rm{i}} e_{\rm{i}} \sim P'_{\rm{i}}/n_{\rm{i}} e_{\rm{i}}\sim 30 \mathrm{kV/m}$, which is beneficial to improve the core plasma confinement \cite{DoyleNF07}. 

In conclusion, we have found that the EIs can generate significant ZFs or mean radial electric fields, by the radial redistribution (RR) of the EI pressure through turbulent transport or external injection or the internal fusion reactions [Eq. (\ref{eq:Er0})]. It is also found that the trapped EIs are more effective in driving ZFs than the isotropic EIs [Eq. (\ref{eq:ErT})]. The new finding suggests that the DT fusion-produced energetic $\alpha$ particles may have significant effects in suppressing turbulence through genearating significant ZFs and mean radial electric fields in a magnetic confinement fusion reactor. 

The RR theory of ZFs presented here is insensitive to the details of the instabilities which cause the RR of EIs. For the electrostatic trubulence, the non-ambipolar effect is given by the poloidal Reynold stress \cite{ZhangNF20}, which denotes the RR of poloidal momentum and is included here. For the RR events caused by the transient electromagnetic instabilities, such as the fishbone or sawtooth bursts, the effect of non-ambipolar RR of particle densities may be neoclassically shielded in contrast to the effect of pressure RR (see, Eq. (\ref{eq:dispersion}) and Refs. \onlinecite{RosenbluthPRL98, WangPoP17}). However, the RR theory presented here may not be applied to the case with steady-state magnetic fluctuations, especially the case with stochastic magnetic field line braiding \cite{RechesterPRL78}; in this case, the anomalous transport induced by the steady-state stochastic magnetic field perturbation is strongly non-ambipolar, and the mean radial electric field is determined by the ambipolarity condition \cite{KaganovichPoP98, YouAIPAdvances23}. 

\begin{acknowledgments}
This work was supported by National Natural Science Foundation under Grant No. 12535014, and the Strategic Priority Research Program of the Chinese Academy of Sciences under Grant Nos. XDB0790201, XDB0500302. 
\end{acknowledgments}

\nocite{*}



\end{document}